\documentclass[fleqn,usenatbib]{mnras}

\usepackage{newtxtext,newtxmath}

\usepackage[T1]{fontenc}

\DeclareRobustCommand{\VAN}[3]{#2}
\let\VANthebibliography\thebibliography
\def\thebibliography{\DeclareRobustCommand{\VAN}[3]{##3}\VANthebibliography}

\usepackage{graphicx}	
\usepackage{amsmath}	

\title[Tiancheng Sun et al.]{Characterising abundance-age relations of GALAH stars using oxygen-enhanced stellar models}

\author[Tiancheng Sun et al.]{
Tiancheng Sun,$^{1,2}$
Xunzhou Chen,$^{3}$\thanks{E-mail: cxz@zhejianglab.com}
Shaolan Bi,$^{1,2}$\thanks{E-mail: bisl@bnu.edu.cn}
Zhishuai Ge,$^{4}$
Maosheng Xiang$^{5,1}$
\newauthor
and Yaqian Wu$^{5}$ 
\\
$^{1}$Institute for Frontiers in Astronomy and Astrophysics, Beijing Normal University,  Beijing 102206, China\\
$^{2}$Department of Astronomy, Beijing Normal University, Beijing 100875, People’s Republic of China\\
$^{3}$Research Center for Intelligent Computing Platforms, Zhejiang Laboratory, Hangzhou 311100, China\\
$^{4}$Beijing Planetarium, Beijing Academy of Science and Technology, Beijing, 100044, China\\
$^{5}$Key Laboratory of Optical Astronomy, National Astronomical Observatories, Chinese Academy of Sciences, A20
Datun Rd., Chaoyang District, \\
Beijing 100101, People’s Republic of China\\
}

\date{Accepted 2023 May 12. Received 2023 May 9; in original form 2023 March 26}

\pubyear{2023}

\begin{document}
\label{firstpage}
\pagerange{\pageref{firstpage}--\pageref{lastpage}}
\maketitle

\begin{abstract}
Main Sequence Turn-off stars (MSTO) and subgiant stars are good tracers of galactic populations. We present a study of 41,034 MSTO and subgiant stars from the GALAH survey. Using a grid of stellar models that accounts for the variation of O abundances, we determine their ages with a median age uncertainty of $\sim$9.4 per cent. Our analysis reveals that the ages of high-O stars based on O-enhanced models (OEM models) are smaller than those determined with $\alpha$-enhanced models, resulting in a mean fractional age difference of $-$5.3 per cent at [O/$\alpha$] = 0.2 and $-$11.0 per cent at [O/$\alpha$] = 0.4. This age difference significantly impacts the age distribution of thick disc and halo stars, leading to a steeper downward trend in the [Fe/H]-age plane from 8 Gyr to 14 Gyr, indicating a shorter formation time-scale and a faster chemical-enhanced history for these populations. We confirm the V-shape of the normalized age–metallicity distribution $p$($\tau$$\mid$[Fe/H]) of thin disc stars, which is presumably a consequence of the second gas infall. Additionally, we find that the halo stars in our sample can be divided into two sequences, a metal-rich sequence (Splash stars) and a metal-poor sequence (accreted stars), with the Splash stars predominantly older than 9 Gyr and the accreted halo stars older than 10 Gyr. Finally, we observe two distinct sequences in the relations between various chemical abundances and age for disc stars, namely a young sequence with ages $<$ $\sim$8 Gyr and an old sequence with ages $>$ $\sim$8 Gyr. 
\end{abstract}

\begin{keywords}
stars: fundamental parameters -- Galaxy: disc -- Galaxy: halo -- Galaxy: evolution -- Galaxy: formation -- Galaxy: abundances
\end{keywords}



\section{Introduction} \label{sec:intro}

The formation and evolution history of the Milky Way is a major challenge in modern astronomy. The Milky Way is typically divided into three primary components, namely the disc, the bulge, and the halo \citep{2016ARA&A..54..529B}. The disc is further classified into a thin disc and a thick disc, which dominate at different heights, and were originally identified based on the observation of a break in the vertical density distribution of stars \citep{1983MNRAS.202.1025G}. It is widely accepted that the thin and thick disc components have different formation histories and timescales, and that stars from each component exhibit distinct chemical, kinematic, and age signatures \citep[e.g.,][]{2012A&A...545A..32A,2018MNRAS.475.5487S}. Various nucleosynthetic production channels exist for different chemical elements, including core-collapse supernovae (CCSNe), white dwarf explosions in binary systems, and asymptotic giant branch stars \citep{2020ApJ...900..179K}. The combination of precise stellar age and chemical abundance enables us to trace the chemical evolution of the Milky Way and provides crucial observational constraints on models of Galaxy formation.

However, estimating stellar ages is a challenging work, and recent efforts have been made to derive more accurate age estimates \citep[e.g.,][]{2017A&A...608A.112N,2019A&A...624A..78D}. Although asteroseismology has achieved significant progress in obtaining precise stellar ages \citep{2018MNRAS.475.5487S}, the sample is largely limited to red giants. The most commonly used method for obtaining stellar ages is grid-based stellar evolution models, such as Y2 isochrones \citep{2001ApJS..136..417Y,2003ApJS..144..259Y,2002ApJS..143..499K,2004ApJS..155..667D}, the Dartmouth Stellar Evolution Database \citep{2008ApJS..178...89D}, and Padova stellar models \citep{2000A&AS..141..371G,2000A&A...361.1023S,2012MNRAS.427..127B}. The chemical composition of heavy elements in these models is a crucial factor that affects age estimation. In theoretical isochrones for modelling metal-poor field stars, $\alpha$-enhanced metal mixtures are commonly employed. In such mixtures, the oxygen abundance is enriched to the same extent as all $\alpha$-elements. Nevertheless, numerous observations over the last two decades have indicated that the oxygen enhancement could be significantly different from other $\alpha$ elements \citep{2005A&A...433..185B, 2006MNRAS.367.1329R, 2014A&A...568A..25N, 2015A&A...576A..89B,2019A&A...630A.104A}. Given that oxygen constitutes a substantial fraction of $\alpha$-elements, it can have a significant influence on stellar evolution and thus age estimation \citep{2007ApJ...666..403D,2012ApJ...755...15V}. Recently, researchers proposed a CO-extreme model to investigate the impact of O enhancements on modelling-inferred masses and ages \citep{2016ApJ...833..161G,2020ApJ...889..157C}. They found that stars with [O/$\alpha$] $>$ 0.2 are globally younger by $\sim$1 Gyr than those determined by the $\alpha$-enhanced metal mixture. Moreover, a recent study using the CO-extreme model determined the ages of 2,926 main-sequence turn-off stars (MSTO) and found that the ages of O-depressed stars are globally older by $\sim$10 per cent compared to those obtained using $\alpha$-enhanced models \citep{2022ApJ...929..124C}. These results demonstrate that variations in oxygen abundance significantly influence the determination of stellar ages.

Main-sequence turn-off (MSTO) and subgiant stars have proven to be valuable tracers of Galactic populations in recent studies \citep{2017RAA....17....5W, 2022ApJ...929..124C, 2022Natur.603..599X}. The ages of MSTO stars can be determined with a high degree of accuracy based on their atmospheric parameters due to the sensitivity of their effective temperature (T$_{\rm eff}$) to their ages at a fixed [Fe/H] \citep{2017RAA....17....5W, 2022ApJ...929..124C}. Subgiant stars exist in a brief phase of stellar evolution during which their luminosity is highly sensitive to their ages, allowing for the most precise and direct age determination \citep{2022Natur.603..599X}. Recent advances in astrometry, such as the accurate parallaxes from the Gaia EDR3 \citep{2021A&A...649A...1G}, have enabled the determination of subgiant star ages with a precision of approximately 7.5 per cent \citep{2022Natur.603..599X}. In this work, we determine the ages of 41,034 MSTO and subgiant stars from GALAH DR3, taking into account the variations in oxygen abundances. Our study aims to investigate the effects of O enhancement on age determination and to present precise abundance-age relations for stars in the Milky Way.

This paper is structured as follows: In Section \ref{sec:data}, we detail our data selection process. In Section \ref{sec:method}, we provide a description of our computational approach for constructing the stellar model grids. We then compare the ages obtained from the O-enhanced models with those obtained from the $\alpha$-enhanced models in Section \ref{sec:age distribution}. The age-abundance relations resulting from our analysis are presented in Sections \ref{sec:subpop1} and \ref{sec:age-chemical}. Finally, in Section \ref{sec:conclusion}, we draw conclusions based on our findings.

\begin{figure*}
\includegraphics[scale=0.65]{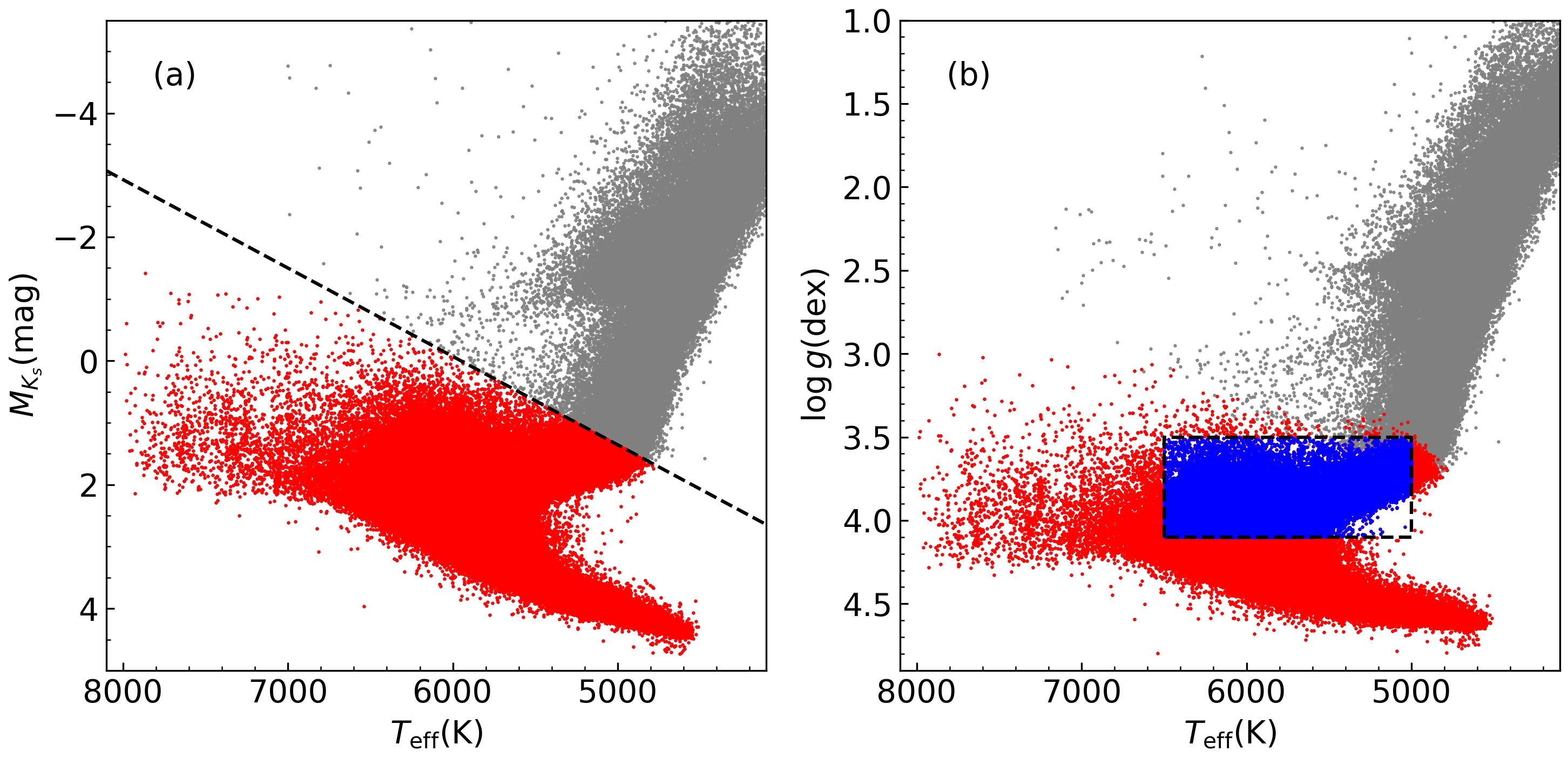}
\caption{ (a) HR diagram of the stars from the GALAH DR3 data (grey dots) and the selected sample (red dots). The black dashed line indicates the cut made to exclude giant stars (see Equation \ref{e0}). (b) Kiel diagram of the stars from the GALAH DR3 data (grey dots), the selected sample (red dots), and the targets used in our work (blue dots). The main-sequence turnoff and subgiant stars are delimited by black dashed lines (3.5 $<\log g<$ 4.1 and 5000 K $<T_{\rm eff}<$ 6500 K). \label{fig:kiel}}
\end{figure*}

\begin{figure*}
\includegraphics[scale=0.58]{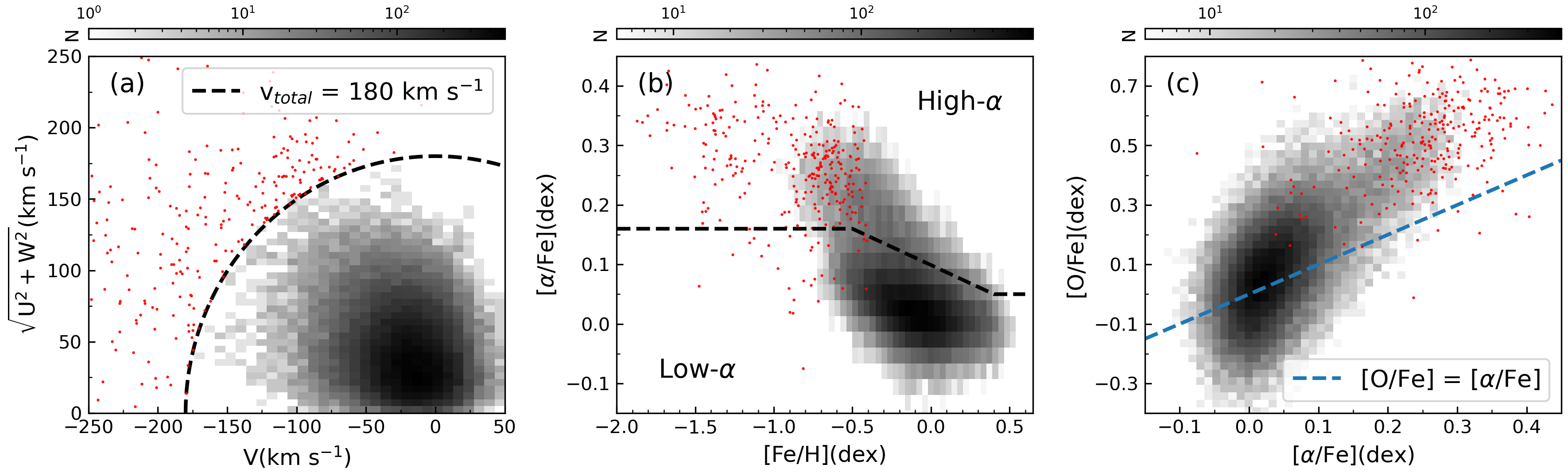}
\caption{Color-coded stellar number density distributions of the targets in the Toomre diagram (a), the [Fe/H]-[$\alpha$/Fe] space (b), and the [$\alpha$/Fe]-[O/Fe] space (c). The black dashed line in panel (a) corresponds to V$_{\rm total}$ = 180 km s$^{-1}$. In panel (b), the high-$\alpha$ and low-$\alpha$ stars are distinguished by whether they lie above or below the three segments of the black dashed line in the [Fe/H]-[$\alpha$/Fe] space, respectively. The blue dashed lines in panel (c) indicate the 1:1 relation. The red dots in each panel represent the halo stars. \label{fig: full sample}}
\end{figure*}

\section{Target selection} \label{sec:data}

GALAH DR3 \citep{2021MNRAS.506..150B} provides stellar parameters ($T_{\rm eff}$, $\log g$, [Fe/H], $V_{mic}$, $V_{broad}$, $V_{rad}$) and up to 30 elemental abundances for 588,571 stars, derived from optical spectra at a typical resolution of R $\sim$ 28,000. Following the recommendations in GALAH DR3, we impose strict selection criteria to ensure reliable stellar parameters, including iron, $\alpha$-elements, and oxygen abundances (flag$\_$sp = 0, flag$\_$fe$\_$h = 0, flag$\_$alpha$\_$fe = 0, and flag$\_$o$\_$fe = 0), requiring an SNR $>$ 30, a chi2$\_$sp $<$ 4 (Chi2 value of stellar parameter fitting), and a quality flag = 0. Binary systems identified by \citet{2020A&A...638A.145T} are excluded. We further impose a single cut based on the Gaia DR3 parameters by selecting stars with a Gaia re-normalized unit weight error (RUWE) less than 1.2. Giant stars are excluded by applying the absolute magnitude cut \citep{2022MNRAS.510.4669S}:
\begin{equation}\label{e0}
M_{K_{s}} =\,m_{K_{s}} - \,A_{K_{s}} - \,5\rm log10[(100\rm\,mas)/\varpi] > 8.5\,-\,T_{\rm eff}/(700\rm\,K)
\end{equation}
Here, the 2MASS $m_{K_{s}}$ magnitudes \citep{2006AJ....131.1163S} and the extinction values $A_{K_{s}}$ are taken from the GALAH catalog. Finally, we select MSTO and subgiant stars satisfying the criteria 3.5 $<\log g<$ 4.1 and 5000 K $<T_{\rm eff}<$ 6500 K, excluding the hottest stars that exhibit temperature-dependent trends in the validation of element abundances \citep{2021MNRAS.506..150B}.

In order to obtain the luminosity of each star, we cross-match our sample with the catalogue from \cite{2023ApJS..264...41Y}. This catalogue provides the luminosity of 1.5 million stars using astrometric data from GAIA DR3 \citep{2022arXiv220800211G} and improved interstellar extinction measurements. Our final sample consisted of 52,261 MSTO and Subgiant stars, which are shown in the Kiel diagram in Figure \ref{fig:kiel}. To study the kinematic properties of halo stars in our sample, we utilized the velocities (U, V, W) and orbital parameters ($L_{Z}$) from the GALAH DR3 value-added catalogue (VAC) \citep{2021MNRAS.506..150B}. These values were derived from the astrometry provided by Gaia EDR3 and radial velocities determined from the GALAH spectra \citep{2021MNRAS.508.4202Z}. The orbital parameters in this catalogue are calculated using the Python package \texttt{Galpy} \citep{2015ApJS..216...29B}, with the details of assumed Milky Way potential and solar kinematic parameters presented in \citet{2021MNRAS.506..150B}.

Figure \ref{fig: full sample} depicts the distribution of the sample stars in the Toomre diagram, the [Fe/H]-[$\alpha$/Fe]\footnote{The [$\alpha$/Fe] values from the GALAH catalogue is calculated as an error-weighted mean of [Mg/Fe], [Si/Fe], [Ca/Fe] and [Ti/Fe].} space, and the [$\alpha$/Fe]-[O/Fe] space. Consistent with the studies by \cite{2010A&A...511L..10N} and \cite{2022MNRAS.510.2407B}, we adopt a criterion to select halo stars by applying a cut in the total velocity of v$_{total}$ $>$ 180 km s$^{-1}$ and [Fe/H] $<$ $-$0.4, as illustrated in Figure \ref{fig: full sample}(a).
Figure \ref{fig: full sample}(b) displays the [Fe/H] versus [$\alpha$/Fe] distribution of sample stars, which exhibits two distinct populations, known as the chemically thin and thick disc stars. As in previous works \citep[e.g.,][]{2012A&A...545A..32A,2014A&A...562A..71B}, we differentiate the high-$\alpha$ stars (thick disc) from the low-$\alpha$ stars (thin disc) using an empirical threshold:

$$ \left\{
\begin{aligned}
& [\alpha/\rm Fe] > 0.16,\;\rm if\;[Fe/H] < -0.5, \\
& [\alpha/\rm Fe] > -0.122 \times \rm [Fe/H] + 0.0989,\;\rm if \\
& -0.5 < [\rm Fe/H] < 0.4, \\
& [\alpha/\rm Fe] > 0.05,\;\rm if\;[Fe/H] > 0.4.   
\end{aligned}
\right.
$$

The diagrams presented in Figure \ref{fig: full sample}(c) depict the relationship between [$\alpha$/Fe] and [O/Fe] for our sample stars. We observe a notable dispersion in [O/Fe] among low-$\alpha$ stars at a given $\alpha$-enhanced value, as indicated by a range from $-$0.4 to +0.6. In contrast, high-$\alpha$ stars exhibit higher [O/Fe] values relative to their [$\alpha$/Fe] values. 


\section{Stellar Models} \label{sec:method}

\begin{figure*}
\includegraphics[scale=0.49]{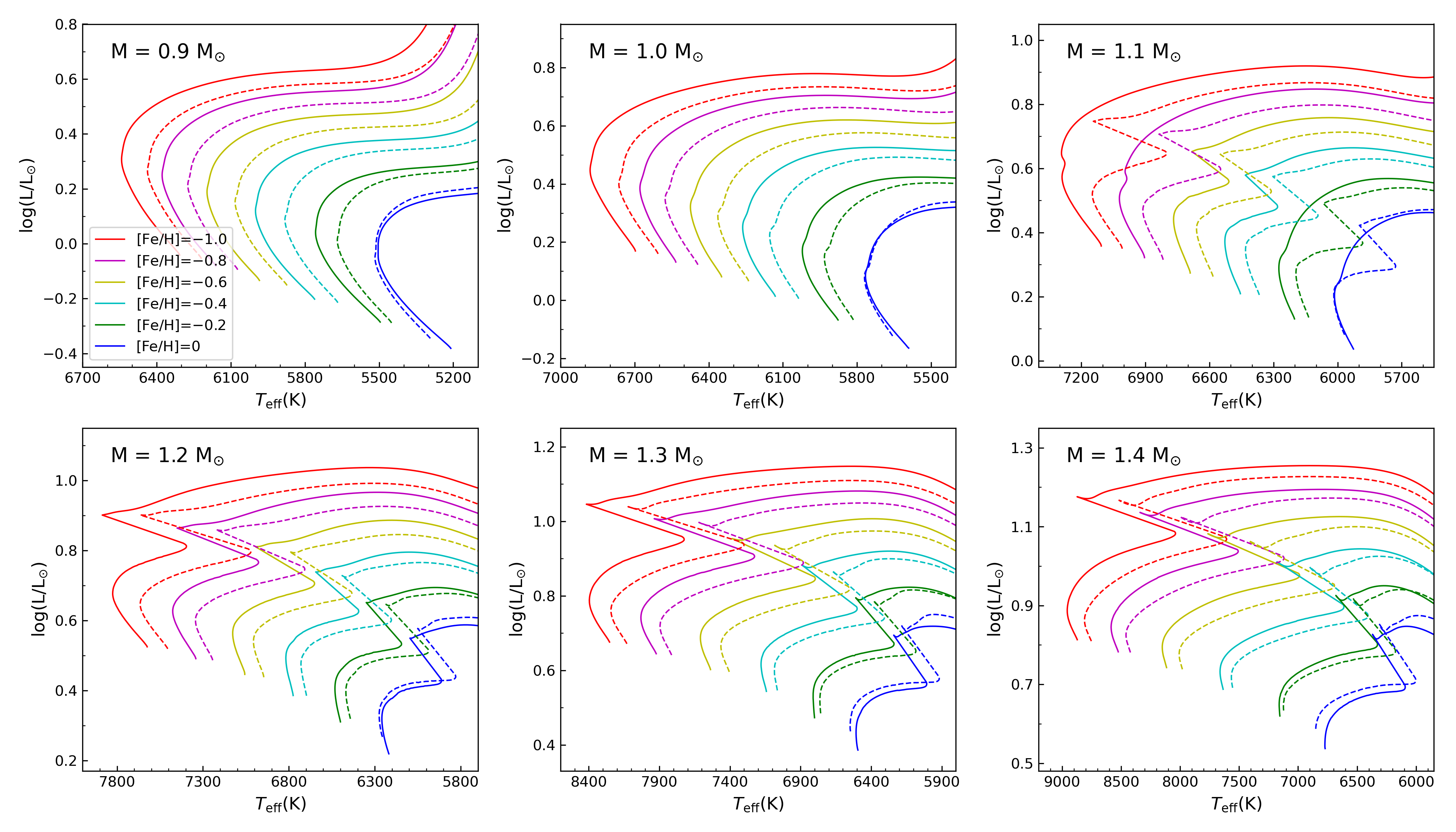}
\caption{Stellar evolution tracks of fixed mass (M = 0.9, 1.0, 1.1, 1.2, 1.3, 1.4 M$_{\odot}$) computed with $\alpha$EM and OEM models. The [Fe/H] range in each panel is 0, $-$0.2, $-$0.4, $-$0.6, $-$0.8, and $-$1.0 (from right to left). The solid and dashed lines represent the tracks with input [O/Fe] = 0.1 and 0.5, respectively. All tracks have the same input [$\alpha$/Fe] (0.1 dex) values. \label{fig:grid}}
\end{figure*}

\begin{figure*}
\includegraphics[scale=0.68]{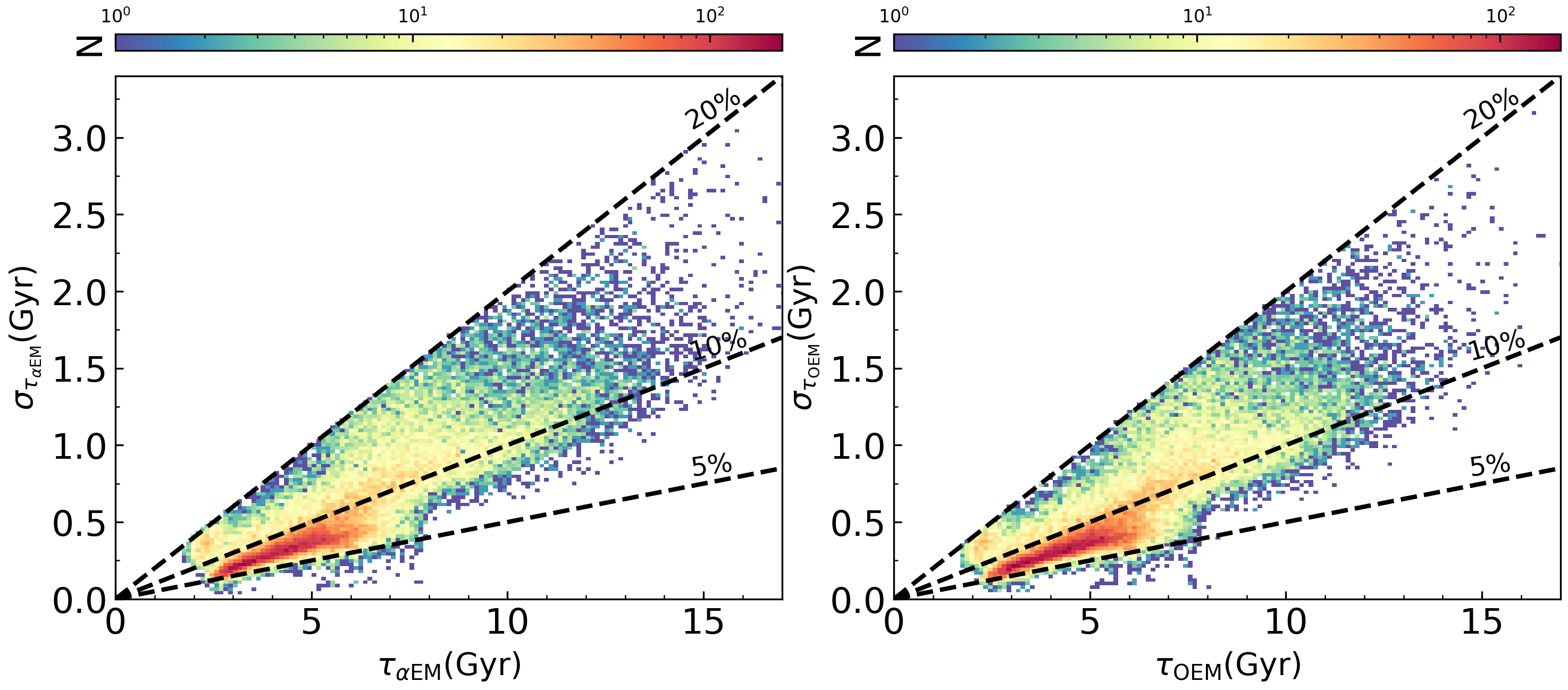}
\caption{Number density distribution in the age uncertainties as a function of age, based on $\alpha$EM models (left) and OEM models (right). Black dashed lines represent the 5 percent, 10 percent and 20 percent fractional uncertainty levels, respectively. \label{fig:age_err}}
\end{figure*}

\begin{figure}
\includegraphics[scale=0.57]{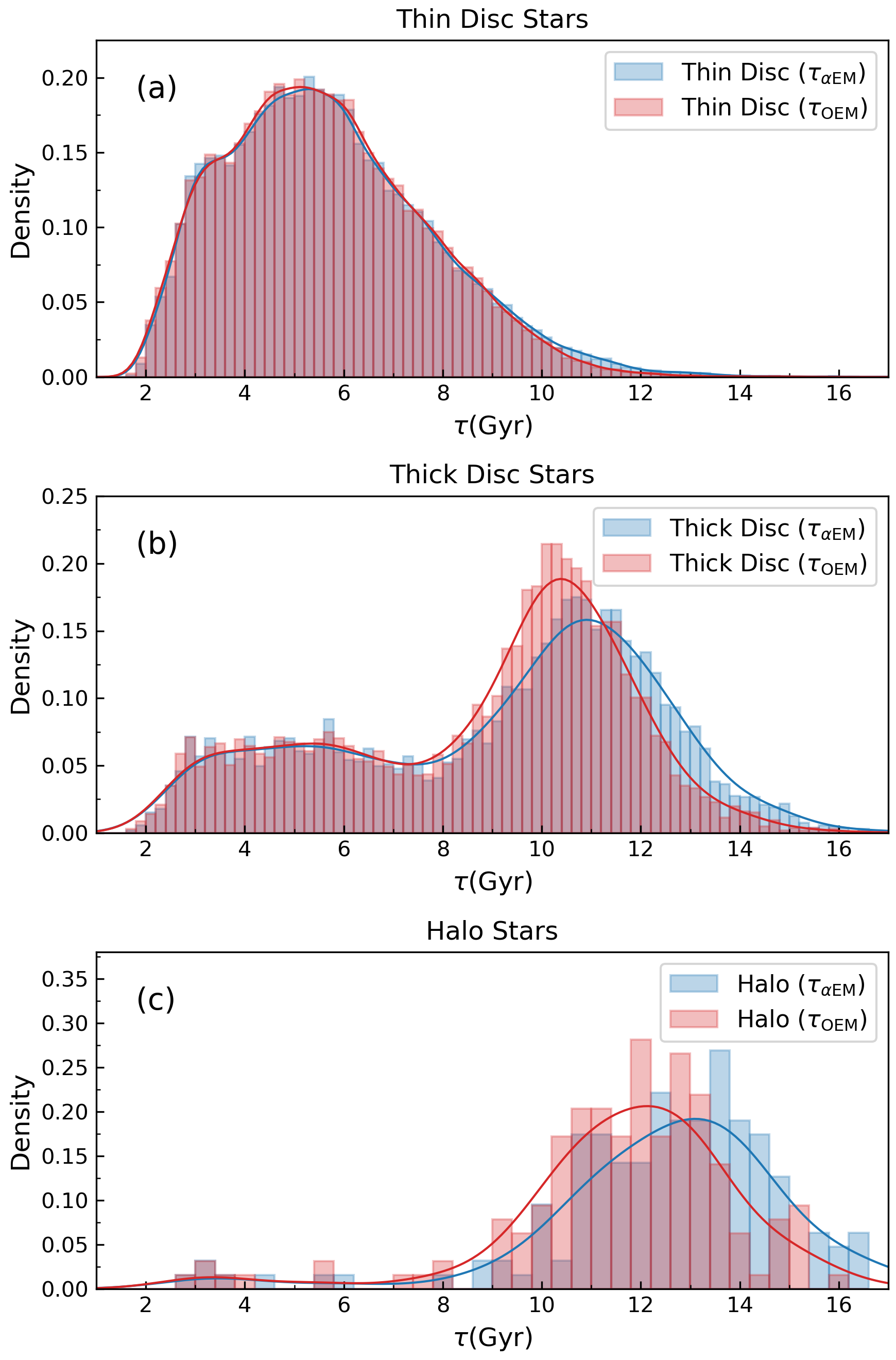}
\caption{Probability density distributions of ages of the thin disc (a), thick disc (b), and halo stars (c). The blue histograms in each panel represent the result from $\alpha$EM models, while the red histograms represent the result from OEM models. The blue curves and red curves represent the kernel density estimates (KDE) of them. \label{fig:age_hist}}
\end{figure}

\begin{figure*}
\centering 
\includegraphics[scale=0.58]{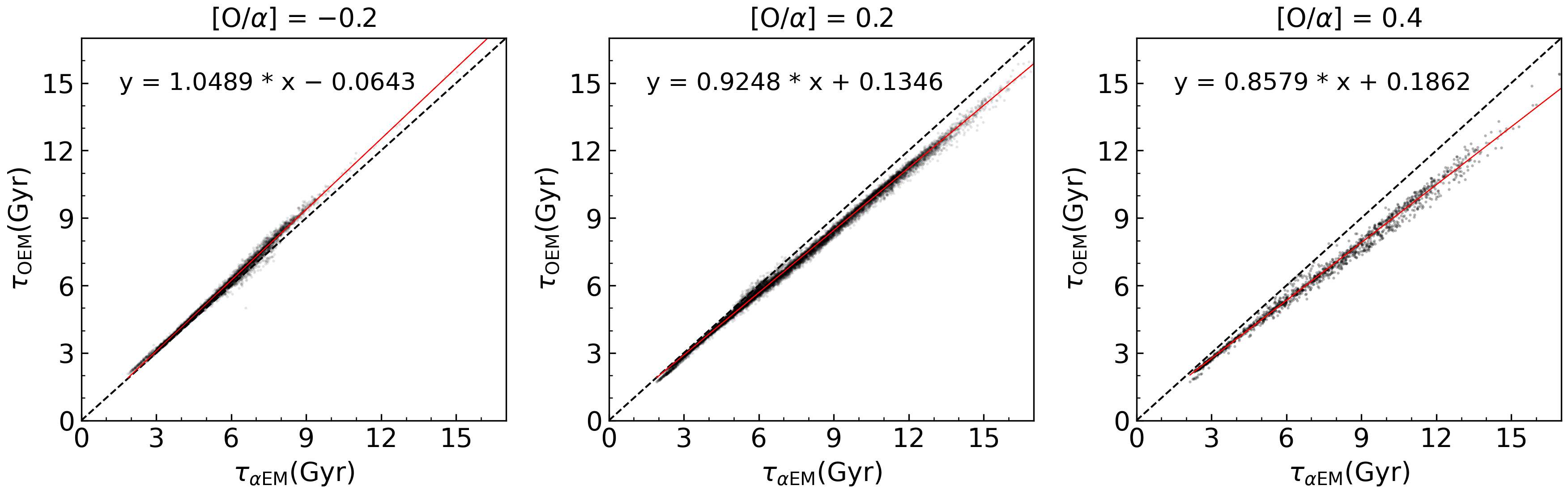} 
\caption{Comparison of the ages determined with $\alpha$EM and OEM models for sample stars. These stars are divided by their [O/$\alpha$] values. Black dash lines show the agonic line. Solid red line indicates the linear fit to the data points, with the fitting result shown in the top-left of each panel.
\label{fig:age_comparison}}
\end{figure*}

\begin{figure*}
\includegraphics[scale=0.58]{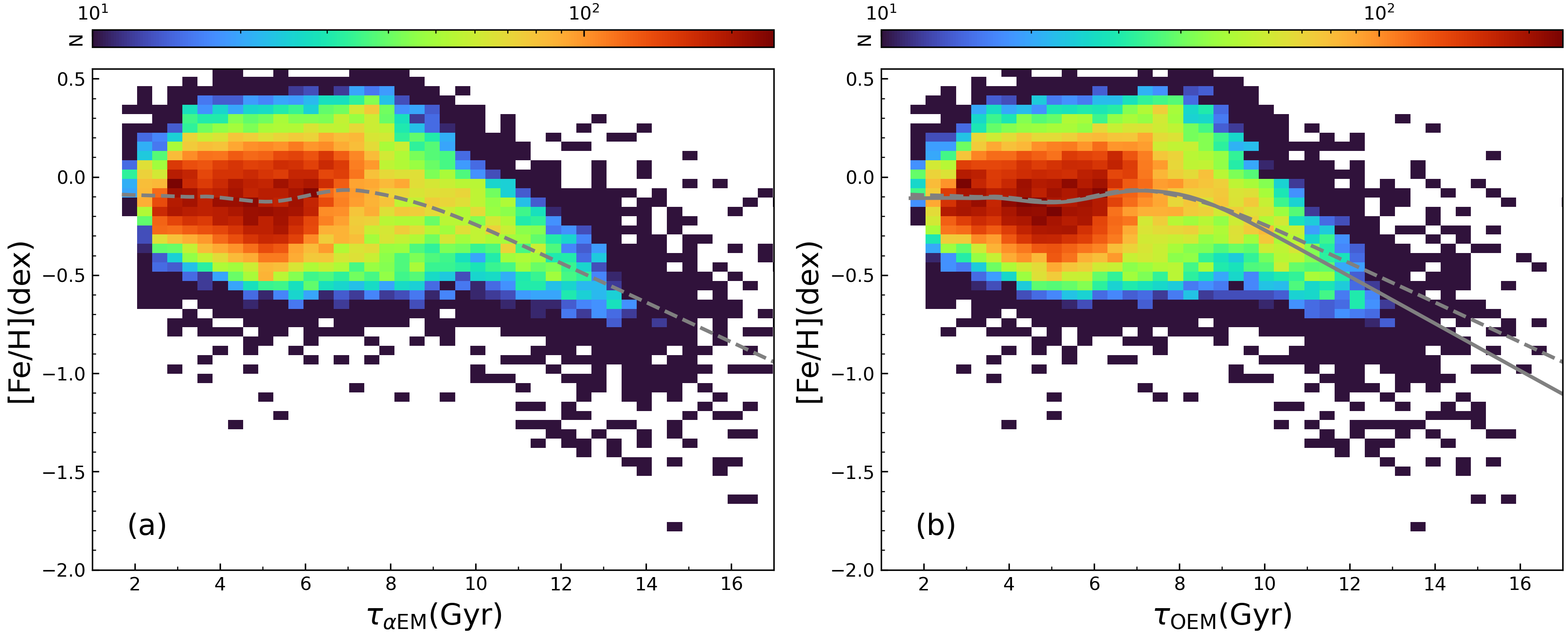}
\caption{Age-[Fe/H] distributions for the full sample, colour-coded by the stellar number density, N. Panel(a) shows the distribution of the full sample based on $\alpha$EM models, with the grey dashed line representing the fitting result by local nonparametric regression. Panel(b) shows the distributions of the full sample based on OEM models. The grey solid line represents the fitting result based on OEM models, and the dashed lines are overplotted for comparison.
\label{fig:age_feh}}
\end{figure*}

\begin{figure*}
\includegraphics[scale=0.58]{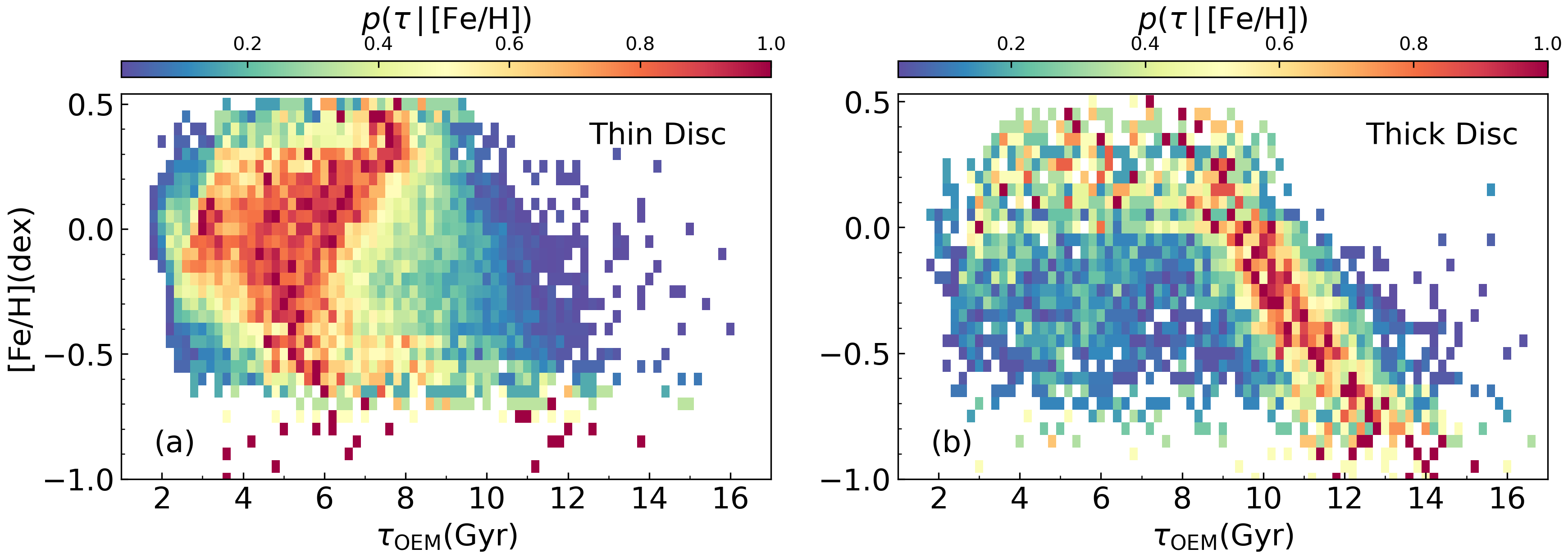}
\caption{Probability distribution of stellar age p($\tau$ | [Fe/H]) (OEM ages), normalized to the peak value for each [Fe/H], for thin disc and thick disc stars.
\label{fig:age_feh_disk}}
\end{figure*}

\begin{figure*}
\includegraphics[scale=0.55]{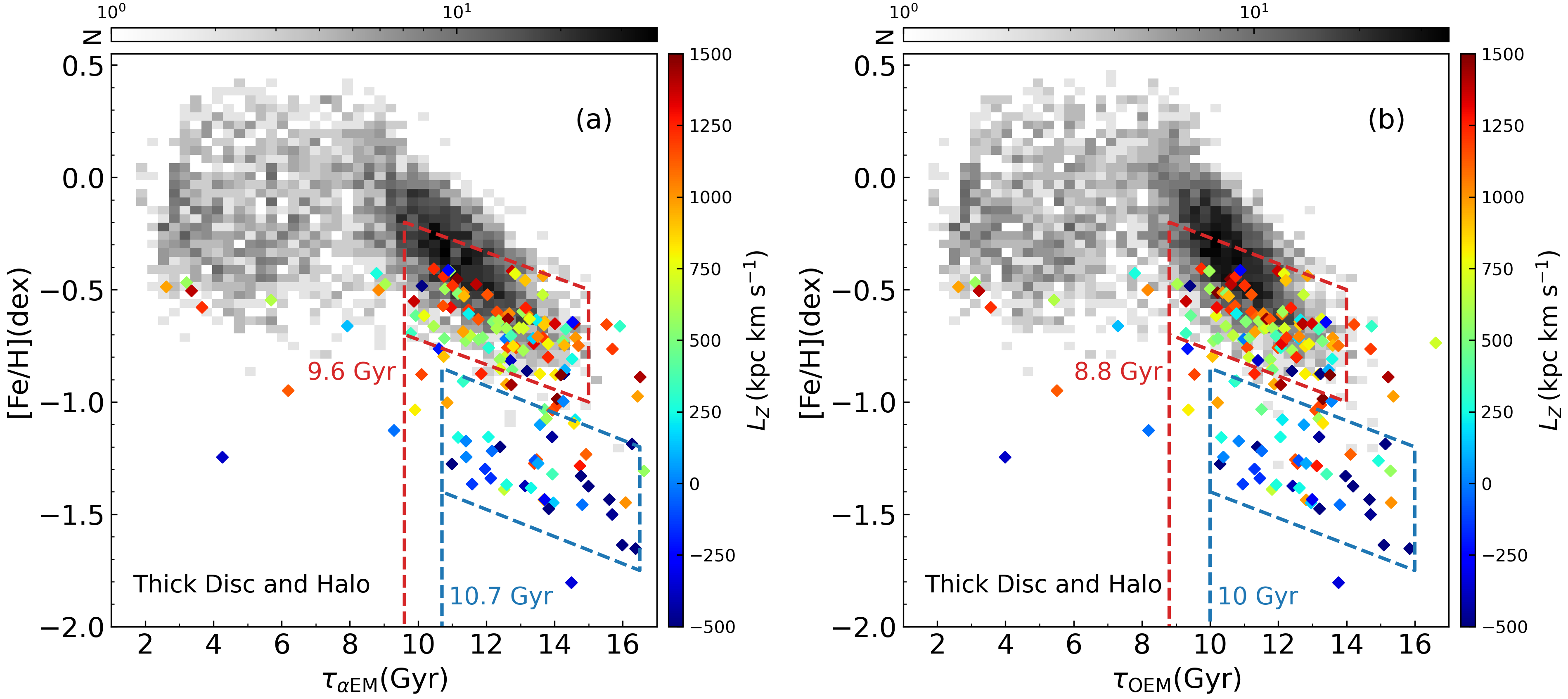}
\caption{Age-[Fe/H] distributions for the thick disc stars (colour-coded by the stellar number density) and halo stars (the overplotted pentagrams). Panel(a) and panel(b) show the results based on $\alpha$EM and OEM models, respectively. The colour of halo stars represents $L_{Z}$. The red and blue dashed boxes in each panel indicate the metal-rich and metal-poor sequences in the halo stars, respectively.
\label{fig:age_feh_halo}}
\end{figure*}

\subsection{Input physics} \label{subsec:Input physics}

\begin{table}
	\centering
	\caption{Grid of Evolutionary Models with Two Metal Mixture Patterns.}
	\label{tbl:table1}
        \setlength{\tabcolsep}{22pt}
	\begin{tabular}{crc} 
		\hline
  		\hline
		Metal-mixture & [O/Fe] & [$\alpha$/Fe] \\
            $ $ & (dex) & (dex) \\
		\hline
            O-enhanced mixture & $-0.2$ & $0$ \\
            $ $ & $0.2$ & $0$ \\
            $ $ & $0.4$ & $0$ \\
            $ $ & $-0.1$ & $0.1$ \\
            $ $ & $0.3$ & $0.1$ \\
            $ $ & $0.5$ & $0.1$ \\
            $ $ & $0.4$ & $0.2$ \\
            $ $ & $0.5$ & $0.3$  \\
            \hline 
            $\alpha$-enhanced mixture & $0$ & $0$ \\
            $ $ & $0.1$ & $0.1$  \\
            $ $ & $0.2$ & $0.2$ \\
            $ $ & $0.3$ & $0.3$  \\
            \hline
	\end{tabular}
\end{table}

We use the Modules for Experiments in Stellar Astrophysics (MESA) code \citep [MESA][]{2011ApJS..192....3P, 2013ApJS..208....4P, 2015ApJS..220...15P, 2018ApJS..234...34P, 2019ApJS..243...10P} to construct a grid of stellar evolutionary models. We utilized MESA Revision 12115 and MESA SDK Version 20.3.1. The MESA equation of state (EOS) employed in this work is a blend of the OPAL \citep{2002ApJ...576.1064R}, SCVH \citep{1995ApJS...99..713S}, PTEH \citep{1995MNRAS.274..964P}, HELM \citep{2000ApJS..126..501T}, and PC \citep{2010CoPP...50...82P} EOSes. Nuclear reaction rates were taken from a combination of NACRE \citep{1999NuPhA.656....3A}, JINA REACLIB \citep{2010ApJS..189..240C}, as well as additional tabulated weak reaction rates \citep{1985ApJ...293....1F, 1994ADNDT..56..231O, 2000NuPhA.673..481L}. The screening effect was included via the prescription of \cite{2007PhRvD..76b5028C}. Thermal neutrino loss rates were taken from \citet{1996ApJS..102..411I}. The helium enrichment law was calibrated with initial abundances of helium and heavy elements of the standard solar model provided by \citet{2011ApJS..192....3P}, resulting in a helium-to-metal enrichment ratio of $Y = 0.248 + 1.3324Z$. We set the mixing-length parameter $\alpha_{\rm MLT}$ to 1.82. In order to account for the effect of microscopic diffusion and gravitational settling of elements in low-mass stars, we employed the formulation of \cite{1994ApJ...421..828T}, which can modify the surface abundances and main-sequence (MS) lifetimes \citep[e.g.,][]{2001ApJ...562..521C,2012MNRAS.427..127B}. We utilized the solar mixture GS98 from \citet{1998SSRv...85..161G} and supplemented the opacity tables with OPAL high-temperature opacities \footnote{\url{http://opalopacity.llnl.gov/new.html}} and low-temperature opacities \citep{2005ApJ...623..585F}.

In line with the approach of \citet{2015MNRAS.447..680G}, we generate metal mixtures by modifying the volume density of elements ($\log N$) based on the GS98 solar mixture. In contrast to the $\alpha$-enhanced metal mixture ($\alpha$EM), we use an individual O enhancement factor, thereby allowing the O abundance to be specified independently. We maintain other $\alpha$-elements (i.e., Ne, Mg, Si, S, Ca, and Ti) with the same enhancement factor.
Based on the observed [O/Fe] and [$\alpha$/Fe] in our sample, we construct a range of opacity tables, which are detailed in Table \ref{tbl:table1}. We refer to mixtures that have different O enhancements relative to the other $\alpha$-elements as O-enhanced mixtures (OEM).

\subsection{Grid computations} \label{subsec:grid}

We establish stellar evolutionary model grids with each [$\alpha$/Fe] and [O/Fe] pairs in Table~\ref{tbl:table1}. The mass range considered is from 0.7 M${\odot}$ to 1.5 M${\odot}$ with a grid step of 0.02 M$_{\odot}$. The input [Fe/H] values vary from $-$2.00 to +0.45 dex with a grid step of 0.05 dex. The computation starts from the Hayashi line and proceeds until the surface gravity $\log g$ reaches 3, covering the evolutionary phases of Main-sequence and Subgiant.

Theoretical considerations suggest that, at a given [Fe/H], variations in [O/Fe] can affect the overall metallicity Z, which is related to opacity and can, in turn, alter the efficiency of energy transfer and the thermal structure of stars. Figure \ref{fig:grid} presents a comparison of evolutionary tracks with different metal mixtures. For [Fe/H] $\leq$ $-$0.2, tracks with O-enhanced mixture (OEM) models exhibit lower $T_{\rm eff}$ and luminosity compared to those with $\alpha$-enhanced metal mixture ($\alpha$EM) models. However, at [Fe/H] = 0, the $T_{\rm eff}$ and luminosity of tracks with OEM models tend to be higher than those with $\alpha$EM models. When comparing tracks of 1.1-1.4 M$_{\odot}$ with those of 0.9-1.0 M$_{\odot}$, we observe that the effects of O enhancement vary with mass. Specifically, for tracks with a mass of 1.1 M$_{\odot}$, a blue hook morphology appears at [Fe/H] $\leq$ $-$0.8, which increases the $T_{\rm eff}$ difference between the two models at this evolutionary phase. At 1.2-1.4 M$_{\odot}$, the models exhibit the same morphology at the end of the main-sequence.

\subsection{Fitting method} \label{subsec:Parameter Estimation}

In this study, we utilize five observed quantities, namely $T_{\rm eff}$, luminosity, [Fe/H], [$\alpha$/Fe], and [O/Fe], to determine fundamental parameters such as stellar mass and age. Notably, the [O/Fe] parameter is excluded in the estimation of parameters using $\alpha$EM models.
 
Following the fitting method introduced by \cite{2010ApJ...710.1596B}, we compare model predictions with their corresponding observational properties $D$ to calculate the overall probability of the model $M_i$ with posterior probability $I$,
\begin{equation}\label{e1}
p\left(M_{i}\mid D,I\right)=\frac{p\left(M_{i}\mid I\right) p\left(D\mid M_{i}, I\right)}{p(D\mid I)}
\end{equation}
where $p$($M_i$ $\mid$ $I$) represents the uniform prior probability for a specific model, and $p$(D $\mid$ $M_i$, $I$) is the likelihood function: 
\begin{equation}\label{e2}
\begin{aligned}
p\left(D\mid M_{i},I\right)=L(T_{eff},[Fe/H],lum)\\
=L_{T_{eff}}L_{[Fe/H]}L_{lum}
\end{aligned}
\end{equation}
The $p$($D$ $\mid$ $I$) in Equation \ref{e1} is a normalization factor for the specific model probability:
\begin{equation}\label{e4}
p(D \mid I)=\sum_{j=1}^{N_{m}} p\left(M_{j} \mid I\right) p\left(D \mid M_{j}, I\right)
\end{equation}
where $N_m$ is the total number of selected models. The uniform priors $p$($M_i$ $\mid$ $I$) can be cancelled, giving the simplified Equation (1) as :
\begin{equation}\label{e5}
p\left(M_{i} \mid D, I\right)=\frac{p\left(D \mid M_{i}, I\right)}{\sum_{j=1}^{N_{m}} p\left(D \mid M_{j}, I\right)}.
\end{equation}
We obtain the probability distribution for each star with Equation \ref{e5} and fit a Gaussian function to the likelihood distribution.
The centre and standard deviation of the Gaussian profile are the estimate and uncertainty, respectively. 
To investigate potential model dependency in age determination, we present a comparison between the results obtained from our $\alpha$EM models and those provided in the GALAH DR3 VAC (see Appendix \ref{Appendix}).

\begin{figure*}
\includegraphics[scale=0.56]{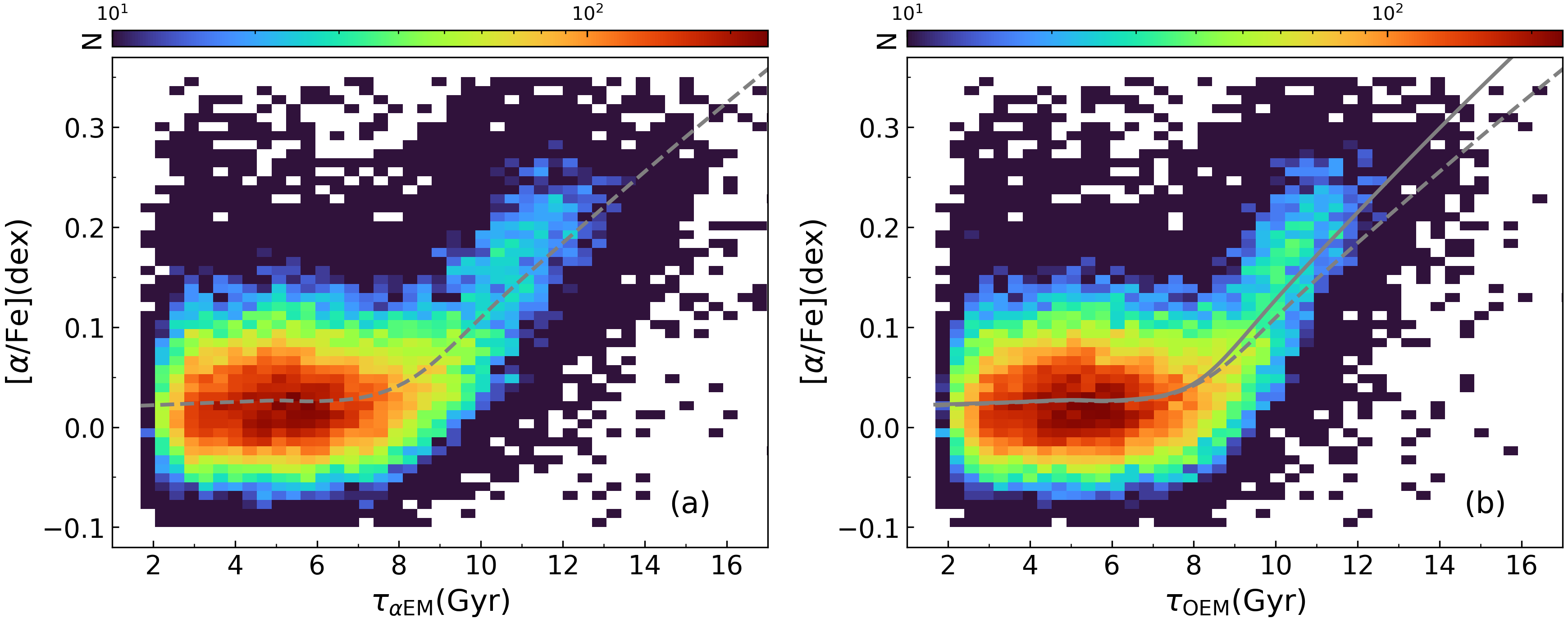}
\caption{Age-[$\alpha$/Fe] distributions for the disc stars. Panel(a) shows the result based on $\alpha$EM models, with the grey dashed line representing the best local nonparametric regression. Panel(b) shows the result based on OEM models. The grey solid line represents the fitting result based on OEM models, and the dashed lines are overplotted for comparison. 
\label{fig:age_alpha}}
\end{figure*}

\begin{figure*}
\includegraphics[scale=0.5]{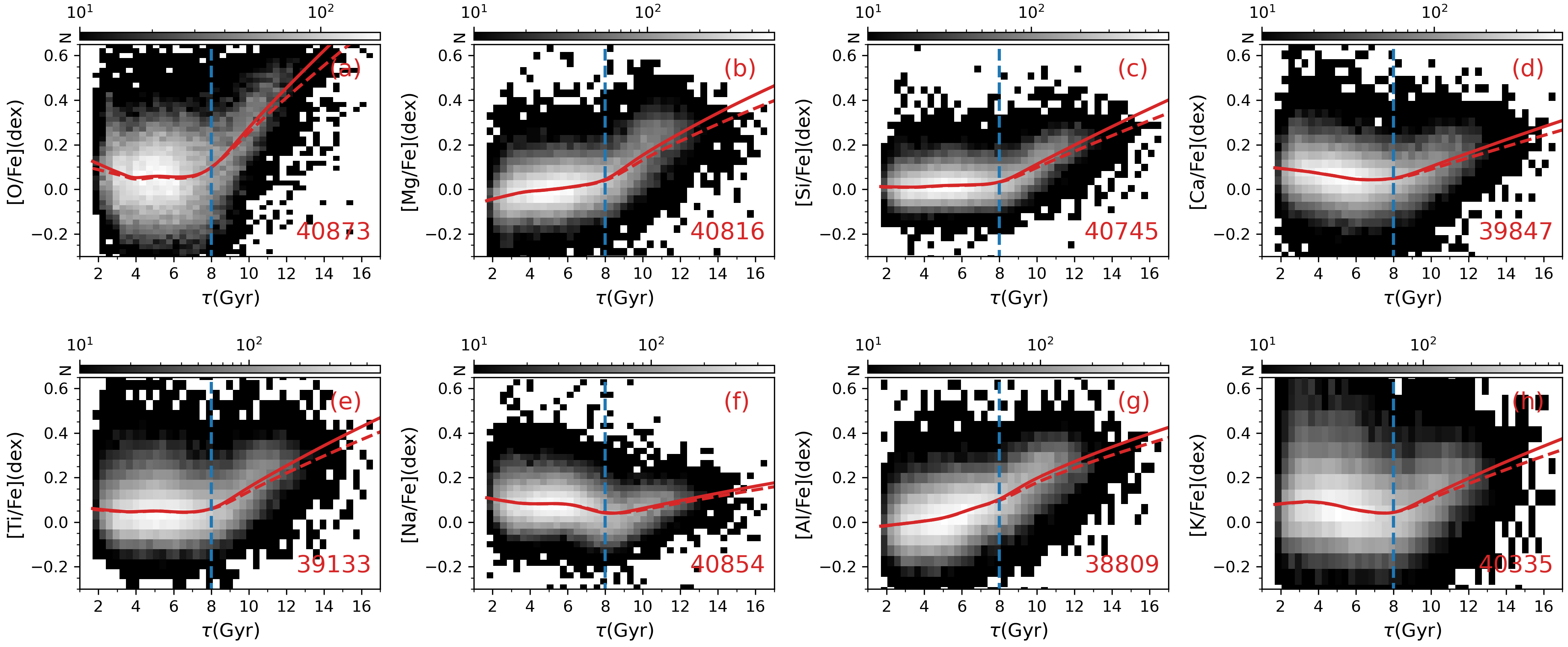}
\caption{Relations between various chemical abundance and ages for the disc stars. The red solid lines in each panel represent the fitting results based on OEM models, and the red dashed lines (results based on $\alpha$EM models) are overplotted for comparison. 
The blue dashed lines in each panel represent the location at age = 8 Gyr. The numbers of stars in each bin are shown in the bottom-right corner of each panel.
\label{fig:age-chemical}}
\end{figure*}

\section{Results} \label{sec:result}

The main objective of this study is to determine the ages of MSTO and subgiant stars considering the variation of O abundance and to investigate their abundance-age relationships. This study provides the masses and ages of sample stars using $\alpha$EM and OEM models. To ensure the accuracy of our results, we remove stars with a relative age uncertainty greater than 20 per cent. Additionally, we exclude 16 stars with significant model systematic bias, whose inferred ages are 2-sigma larger than the age of the universe (13.8 Gyr) according to \citet{2016A&A...594A..13P}. After applying these cuts, we obtain ages for 41,034 GALAH MSTO and subgiant stars.

The age uncertainties for our sample are presented in Figure \ref{fig:age_err}, where we compare the uncertainties obtained using $\alpha$EM (($\tau_{\alpha\rm EM}$)) and OEM models ($\tau_{\rm OEM}$). Our sample stars have a median relative age uncertainty of 9.4 per cent across the age range of $\sim$1.5 Gyr to 13.8 Gyr.

\subsection{Stellar ages} \label{sec:age distribution}

We present the age distributions of the thin disc, thick disc, and halo stars, as shown in Figure \ref{fig:age_hist}. 
The age distribution of the thin disc (Figure \ref{fig:age_hist}(a)) exhibits two clear peaks, which is also found in \citet{2022MNRAS.510.4669S}. Specifically, the distribution displays a young peak at 3.5 Gyr ($\alpha$EM models) or 3.3 Gyr (OEM models), and an old peak at 5.3 Gyr ($\alpha$EM models) or 5.1 Gyr (OEM models). The thick disc stars in Figure \ref{fig:age_hist}(b) have a peak age at 10.7 Gyr ($\alpha$EM models) or 10.2 Gyr (OEM models) with a younger tail reaching $\sim$1.5 Gyr. Some of the young thick disc stars in our sample may be due to contamination from thin disc stars, and there are also young-$\alpha$ rich stars attributed to the binary evolution of old stars in previous works \citep{2015MNRAS.451.2230M,2018MNRAS.475.5487S,2019MNRAS.487.4343H,2020ApJ...903...12S,2021ApJ...922..145Z}. Compared to the results from $\alpha$EM model, the thick disc stars' ages are significantly younger, particularly for stars older than 9 Gyr, indicating a shorter formation time-scale for the thick disc. Figure \ref{fig:age_hist}(c) displays the age distributions of halo stars, which have a median age of 12.5 Gyr using $\alpha$EM models and 11.7 Gyr with OEM models. The peak ages of halo stars and thick disc stars from the OEM model are much closer than those from the $\alpha$EM model, and the ages from the OEM model are more reasonable, as there are fewer values beyond the age of the universe (13.8 Gyr).

Figure \ref{fig:age_comparison} presents a comparison between ages determined using $\alpha$EM models and OEM models. The sample stars are divided into two groups based on their [O/$\alpha$] values. High-O stars, with [O/$\alpha$] $>$ 0, and low-O stars, with [O/$\alpha$] $<$ 0, are referred to as such throughout the rest of this paper. Based on the OEM models, high-O stars have younger ages than those determined by $\alpha$EM models, while low-O stars have older ages. Moreover, there exists a significant correlation between the age offsets and [O/$\alpha$] values. Specifically, at [O/$\alpha$] = $-$0.2, low-O stars have a mean fractional age difference of 3.7 per cent, whereas at [O/$\alpha$] = 0.2 and 0.4, high-O stars have mean fractional age differences of $-$5.3 per cent and $-$11.0 per cent, respectively.

\subsection{Age–metallicity distributions} \label{sec:subpop1}

Figure \ref{fig:age_feh} illustrates the age–[Fe/H] relation of the sample stars. Local nonparametric regression fitting (LOESS model) is applied to the data in each panel. Figure \ref{fig:age_feh}(a) and Figure \ref{fig:age_feh}(b) depict the age–[Fe/H] distributions of all stars based on $\alpha$EM models and OEM models, respectively. The result reveals an evident increasing trend in [Fe/H] with age from 14 Gyr to $\sim$8 Gyr and a mostly flat trend with large dispersion for ages $<$ 8 Gyr. Notably, the OEM models show a steeper trend with age at the early phase of the Milky Way (from 14 Gyr to 8 Gyr) compared to the $\alpha$EM models, which implies a shorter formation timescale of the old population.

In order to examine the age-[Fe/H] distributions of thin and thick disc stars, we employ a normalization procedure for the distribution $p$($\tau$,[Fe/H]) to obtain $p$($\tau$$\mid$[Fe/H]), the age distribution at a specified [Fe/H]. By normalizing the distribution at each [Fe/H], we can clearly discern the trend of [Fe/H] versus age, without being affected by the number of stars in each [Fe/H] bin. As shown in Figure \ref{fig:age_feh_disk}(a), the thin disc stars exhibit a clear V-shape \citep{2018MNRAS.477.2326F,2022Natur.603..599X}, with a metal-rich branch ([Fe/H] $\gtrsim$ $-$0.1) stretching from [Fe/H] $\simeq$ 0.5 at 8 Gyr to [Fe/H] $\simeq$ $-$0.1 at 4-6 Gyr, and a metal-poor branch ([Fe/H] $\lesssim$ $-$0.1) raising its [Fe/H] from $\simeq$ $-$0.7 at 8 Gyr to [Fe/H] $\simeq$ $-$0.1 at 4-6 Gyr. 
Another interesting feature in Figure \ref{fig:age_feh_disk}(a) is an overdensity at an age of around 3 Gyr, which occurs subsequent to the convergence of stars in two branches towards [Fe/H] $\simeq$ $-$0.1 at 4-6 Gyr.
As for the thick disc stars, their distributions in Figure \ref{fig:age_feh_disk}(b) reveal a clear and tight age–metallicity relation from [Fe/H] $\simeq$ $-$1 at 14 Gyr to [Fe/H] $\simeq$ 0.5 at 8 Gyr ago, which is consistent with the findings of previous works \citep{2013A&A...560A.109H}.

The V-shaped distribution observed in thin disc stars can be explained by the revised "two-infall" chemical evolution model \citep{2017MNRAS.472.3637G, 2019A&A...623A..60S,2020MNRAS.498.1710P}. According to this model, the thick disc is formed by a gas infall episode, followed by the formation of the thin disc over a longer timescale through an independent gas accretion event. Our results suggest that the metal-poor gas from the second accretion event gradually dominates the later star formation since $\sim$8 Gyr ago, resulting in the declining trend of the metal-rich branch observed in Figure \ref{fig:age_feh_disk}(a).
The overdensity at $\sim$3 Gyr is related to the young peak of thin disc in Figure \ref{fig:age_hist}(a), which is consistent with previous works \citep{2019A&A...624L...1M,2019ApJ...878L..11I,2022MNRAS.510.4669S}, indicating a recent burst of star formation $\sim$3 Gyr ago in the disc \citep{2021MNRAS.508.4484J}.

We examine the age-[Fe/H] distributions of halo stars in our sample and analyze their angular momentum ($L_{Z}$) to classify them into two sequences, as shown by the red dashed box and blue dashed box in Figure \ref{fig:age_feh_halo}. For comparison, we present the number density distribution of thick disc stars in each panel. The halo stars belonging to the metal-rich sequence (inside the red box) have larger $L_{Z}$ values (peak value of $\sim$500 kpc km s$^{-1}$) and are located near the oldest and most metal-poor part of the thick disc stars in the age-[Fe/H] plane. The disc-like age-[Fe/H] distribution of the metal-rich sequence suggests that these stars were formed in situ within the Galactic disc and were later ‘splashed’ to relatively low-angular-momentum orbits (compared to the typical angular momentum of disc stars) with a halo-like total velocity during an early merger event, i.e., the merger with the Gaia-Enceladus/Sausage satellite galaxy \citep{2017ApJ...845..101B,2020MNRAS.494.3880B}.
The metal-poor sequence (inside the blue box) is primarily composed of stars with $L_{Z}$ values ranging from $-$500 to 500 kpc km s$^{-1}$, suggesting that they may have originated from the satellite galaxy and are therefore considered accreted stars. Our OME model-based results show that the ages of stars in both sequences are approximately 1 Gyr younger than those predicted by $\alpha$EM models. Our result indicates that the majority of Splash stars have ages $\gtrsim$ 9 Gyr, which is slightly smaller than the minimum age of 9.5 Gyr reported in previous studies \citep{2020MNRAS.494.3880B}. However, the majority of accreted halo stars in our sample are older than 10 Gyr, which agrees with previous studies regarding the satellite galaxy Gaia-Enceladus/Sausage \citep{2018ApJ...860L..11K,2018Natur.563...85H,2021NatAs...5..640M,2022MNRAS.510.4669S}.

\subsection{[X/Fe] versus stellar age} \label{sec:age-chemical}
It is generally understood that $\alpha$-elements are primarily produced by short-lived stars via winds and core-collapse supernovae (CCSNe), while Fe is predominantly produced by Type Ia SNe with only a small fraction being ejected by CCSNe. The longer timescales for Type Ia SNe explosions, in contrast to those of CCSNe, make ratios such as [$\alpha$/Fe] useful as cosmic clocks \citep{2013A&A...560A.109H,2014A&A...562A..71B,2017A&A...608L...1H}. In this context, we present the age–[$\alpha$/Fe] distributions of the stars in our sample in Figure \ref{fig:age_alpha}.

Our findings demonstrate a strong correlation between age and [$\alpha$/Fe] in disc stars, where a nearly flat trend is observed in the thin disc population between the ages of 2 Gyr and 8 Gyr, and an increase in [$\alpha$/Fe] with age is evident after 8 Gyr. These results confirm that the thin and thick disc populations have distinct chemical evolution histories \citep{2013A&A...560A.109H,2018MNRAS.475.5487S,2020A&A...640A..81N}. By comparing the results from $\alpha$EM and OEM models, we observe a steeper rising trend in the thick disc population between the ages of 8 Gyr and 14 Gyr, suggesting a more rapid chemical enrichment history, consistent with Figure \ref{fig:age_feh}. We then investigate the age-[X/Fe] relations for individual $\alpha$ elements (O, Mg, Si, Ca, Ti) and odd-Z elements (Na, Al, K), which are mainly produced by CCSNe \citep{2020ApJ...900..179K}. As depicted in Figure \ref{fig:age-chemical}, except for Al, most elements exhibit two trends: a young sequence with ages $<$ $\sim$8 Gyr and an old sequence with ages $>$ $\sim$8 Gyr. OEM modelling shows that the abundances of these elements experience a sharper increase at ages greater than 8 Gyr, indicating a higher formation rate for the old sequence of disc stars.

\section{Conclusions}
\label{sec:conclusion}

To determine the ages of MSTO and subgiant stars with the variation of O abundance and investigate their abundance-age relations, we constructed a grid of stellar models that consider oxygen abundance as an independent input. We derived the masses and ages of 41,034 GALAH stars and performed an extensive analysis of their chemical properties combined with the ages obtained from both $\alpha$-enhanced and O-enhanced models.

Our main conclusions are summarized as follows:

(\romannumeral1) The ages of high-O stars determined with O-enhanced models are generally smaller than those obtained with $\alpha$-enhanced models, while the ages of low-O stars are relatively older. Specifically, our results reveal a mean fractional age difference of 3.7 per cent for low-O stars at [O/$\alpha$] = $-$0.2, whereas high-O stars exhibit mean fractional age differences of $-$5.3 per cent at [O/$\alpha$] = 0.2 and $-$11.0 per cent at [O/$\alpha$] = 0.4.

(\romannumeral2) The ages of the thick disc and halo stars calculated with the OEM models exhibit a significant difference from those derived from the $\alpha$EM models. According to the OEM models, the thick disc stars exhibit a peak at approximately 10.2 Gyr, which is younger than the age peak derived from $\alpha$EM models (10.7 Gyr). On the other hand, the median age of halo stars is 12.5 Gyr based on $\alpha$EM models, whereas OEM models suggest a median age of 11.7 Gyr. The peak ages of the halo stars and thick disc stars estimated from OEM models are much closer to those obtained from $\alpha$EM models.

(\romannumeral3) For all stars in our sample, we find that the trend of [Fe/H] with age based on the OEM model from 8 Gyr to 14 Gyr is steeper compared to the result from $\alpha$EM models. This suggests a higher formation rate for the old sequence, which consists mainly of thick disc stars, and a faster chemical-enhanced history. The age–metallicity distribution $p$($\tau$$\mid$[Fe/H]) of thin disc stars displays a V-shape, with a metal-rich branch ([Fe/H] $\gtrsim$ $-$0.1) and a metal-poor branch ([Fe/H] $\lesssim$ $-$0.1), which is likely due to the second gas infall.

(\romannumeral4) Our study confirms the presence of two distinct sequences in the halo stars: a metal-poor sequence known as Splash stars, and a metal-rich sequence consisting of accreted stars. We find that the ages of stars in both sequences, based on OME models, are younger by approximately 1 Gyr compared to the results obtained using $\alpha$EM models. Our analysis suggests that the majority of Splash stars are older than 9 Gyr, while the accreted halo stars are older than 10 Gyr.

(\romannumeral5) Our analysis reveals a pronounced and steep increasing trend of individual $\alpha$ elements (O, Mg, Si, Ca, Ti) and odd-Z elements (Na, Al, K) with age based on the OEM models, covering the age range from 8 Gyr to 14 Gyr. Moreover, we observe that the abundance-age relations of these elements (except Al) exhibit two sequences: a young sequence characterized by ages $<$ $\sim$8 Gyr, and an old sequence corresponding to ages $>$ $\sim$8 Gyr.

\section*{Acknowledgements}

We thank the referee and scientific editor for helpful comments and suggestions that improved the presentation of the manuscript.
We thank Ferguson J. W. for providing low-temperature opacities utilized in this work.
This work used the data from the GALAH survey, which is based on observations made at the Anglo Australian Telescope, under programs A/2013B/13, A/2014A/25, A/2015A/19, A/2017A/18, and 2020B/23.
This work has made use of data from the European Space Agency (ESA) mission Gaia (\url{https://www.cosmos.esa.int/gaia}), processed by the Gaia Data Processing and Analysis Consortium (DPAC, \url{https://www.cosmos.esa.int/web/gaia/dpac/consortium}). Funding for the DPAC has been provided by national institutions, in particular the institutions participating in the Gaia Multilateral Agreement.
This work is supported by National Key R$\&$D Program of China No. 2019YFA0405503, the Joint Research Fund in Astronomy (U2031203) under cooperative agreement between the National Natural Science Foundation of China (NSFC) and Chinese Academy of Sciences (CAS), and the NSFC grants (12090040, 12090042). This work is partially supported by the Scholar Program of Beijing Academy of Science and Technology (DZ:BS202002).

\section*{Data Availability}
 
The data underlying this article will be shared on reasonable request to the corresponding author.

\appendix

\section{IMPACT OF STELLAR MODELS ON AGE DETERMINATION}
\label{Appendix}

\begin{figure}
\includegraphics[scale=0.55]{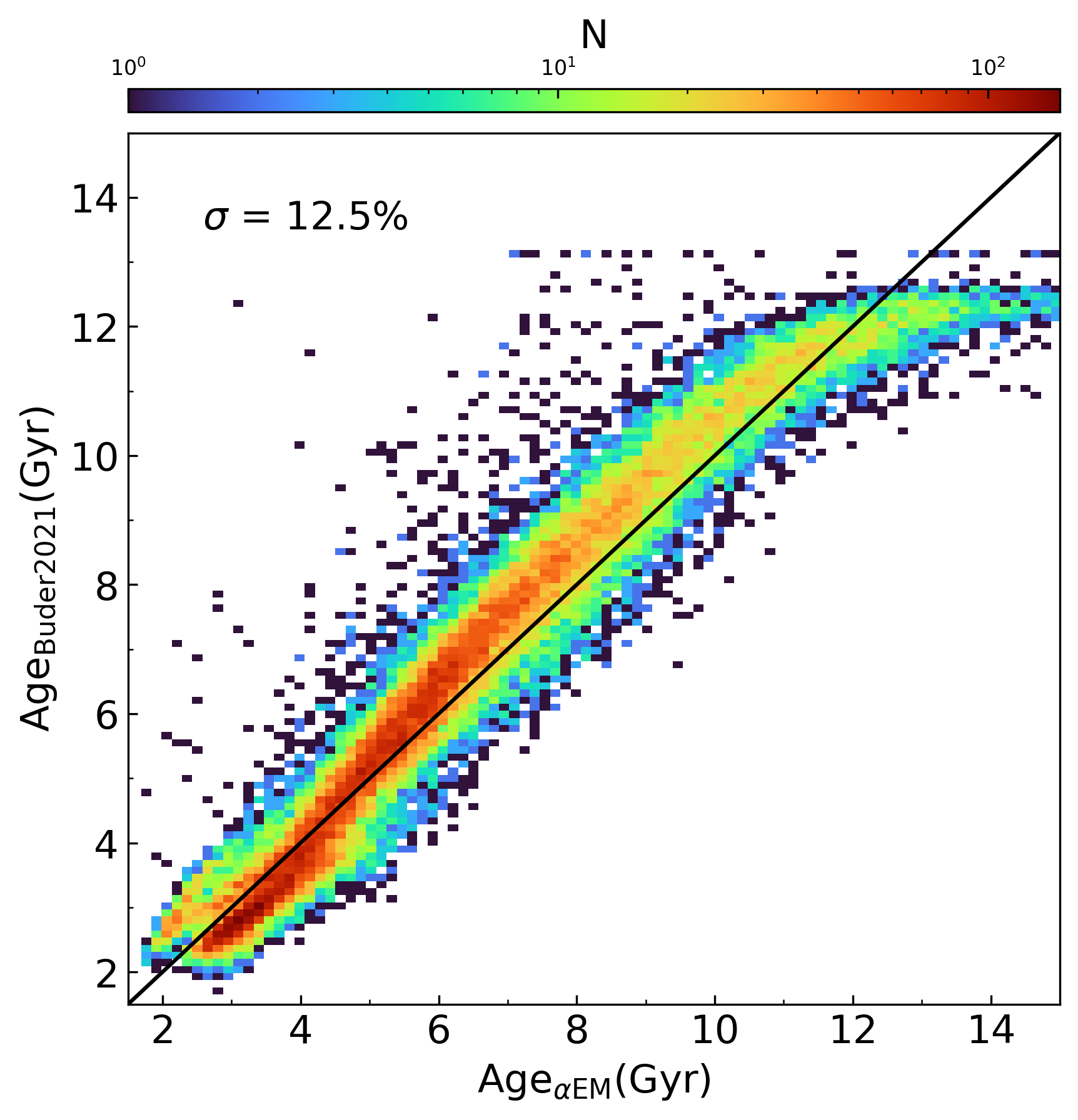}
\caption{Comparison of ages of 35,776 GALAH sample stars from our $\alpha$EM models and the GALAH DR3 value-added catalogue \citep[VAC,][]{2021MNRAS.506..150B}. The black line represents the 1:1 line. \label{fig:age_com_vac}}
\end{figure}

Figure \ref{fig:age_com_vac} shows the comparison of age estimations for 35,776 GALAH stars, with a relative age uncertainty of less than 20\%, based on $\alpha$EM models and GALAH DR3 VAC \citep{2021MNRAS.506..150B}.
The ages of stars from GALAH DR3 VAC are calculated using the PARSEC (the PAdova and TRieste Stellar Evolution Code) release v1.2S + COLIBRI stellar isochrone \citep{2017ApJ...835...77M}, which adopt a solar-scaled metal mixture, i.e., input [$\alpha$/Fe] = 0. 
Figure \ref{fig:age_com_vac} illustrates that the one-to-one relation of the results is quite good for most stars. However, it is noteworthy that the adopted approach in GALAH DR3 VAC includes a flat prior on age with an age cap of 13.2 Gyr \citep{2018MNRAS.473.2004S}. Consequently, the ages of the majority of stars from GALAH DR3 VAC are found to be younger than 12 Gyr, which results in a relatively large dispersion of age differences (12.5\%).
In addition, we observe systematic differences between the PARSEC and $\alpha$EM models, with the PARSEC model yielding age estimates that are 2.8\% older than those obtained from our $\alpha$EM models. These discrepancies can be attributed to differences in the input physics employed by the two models, such as the input [$\alpha$/Fe] value, helium abundance, and mixing-length parameter.



\bibliographystyle{mnras}
\bibliography{ref} 

\bsp	
\label{lastpage}
\end{document}